\documentclass[12pt,a4paper]{article}
\usepackage{graphicx}
\usepackage{setspace}
\usepackage{natbib}
\usepackage{amssymb}
\usepackage{epsfig,color}
\usepackage{fancyhdr}

\begin{document}
% \title{Mass functions of clumps}
% \maketitle
% \makebox[3\width]{%
% A poster report, presented at the conference}
% \par
% \makebox[1\width]{%
% {\it Galactic Scale Star Formation: Observation meets theory,} Heidelberg, July 30 -August 3, 2012
% } \par

\pagestyle{plain}
% \setcounter{page}{1}
% \pagenumbering{arabic}

\flushright{
{\bf \emph{A poster report, presented at the conference\\
Galactic Scale Star Formation: Observation meets theory,\\
Heidelberg, July 30 -August 3, 2012}}
}
\vspace*{24pt}
\begin{center}
{\LARGE \bf Modeling mass functions of clumps\\ formed during the early MC evolution\vspace*{18pt}\\}
{\Large Todor V. Veltchev$^{\,1,\,2}$, Sava Donkov$^3$ \vspace*{10pt}}
\flushleft{\large \it
$^1$ Faculty of Physics, University of Sofia, Bulgaria \\
$^2$ Institute of Theoretical Astrophysics, Heidelberg, Germany\\
$^3$ Department of Applied Physics, Technical University, Sofia, Bulgaria \vspace*{6pt}\\}
{\footnotesize E-mails:~eirene@phys.uni-sofia.bg,~savadd@tu-sofia.bg \vspace{12pt}\\}
\end{center}

\begin{quote}
{\bf Abstract:} The statistical approach for description of molecular cloud substructure, proposed by Donkov, Veltchev and Klessen (2011, 2012), allows for alternative models, operating with different type of objects: an ensemble of clumps or a larger cloudlet. We demonstrate briefly the predictive power of both models, applied to molecular emission and dust extinction studies of Galactic clouds.
\end{quote}

\flushleft

\section{Physical description of clumps}
Clumpy, dense structures in molecular clouds (MCs) and/or in the cold neutral medium are direct progenitors of protostars or of protostellar clusters. While numerous algorithms for identification of coherent regions on column-density maps or in the position-velocity space have been suggested \citep[e.g.,][]{SG_90, WdGB_94, Mensh_ea_10}, the issue whether and how those objects represent the existing clumps is open. Therefore it is useful to invent a {\it statistical approach} for their description \citep{DVK_11, DVK_12}.\vspace*{12pt}\\ 

A physically sound basis for such approach is to consider clumps as condensations of turbulent (shock) origin formed and shaped during the early MC evolution while gravity slowly takes over at small spatial scales. Assuming that compact (quasi-spherical or cubic) clumps with sizes $l$, significantly lower than the scale of their generation $L$, and of various densities $\rho$ in respect to the local mean density $\langle \rho \rangle_L$ populate the whole considered volume $L^3$, we adopt a natural physical relation for their masses:
\[ \frac{m}{m_0}=\frac{\rho}{\langle \rho \rangle_L}\,\Bigg(\frac{l}{l_0}\Bigg)^3~,\]
where $m_0$ and $l_0$ are units of normalization. The building blocks of the model stem from the physical consideration of: i) isothermal, isotropic, fully saturated turbulence in a steady state; and, ii) the role of gravity for the energy balance at different spatial scales.
\newpage
\section{Basic assumptions of our statistical approach}
  \begin{itemize}
\item[1)] ~Turbulent cascade in isothermal medium in the scale range $0.5\lesssim L \lesssim 20$~pc, conditioning scaling laws of velocity and density according to \citet{Larson_81}:
\begin{eqnarray}
	u & \propto & L^{\beta}~,~~~~~~~~0.33\lesssim\beta\lesssim0.65 \nonumber\\ 
 	\langle\rho\rangle & \propto & L^{\alpha}           \nonumber	 
	\end{eqnarray}
 \item[2)] ~Lognormal probability density distribution (PDF) at each scale $L$ with stddev 
\[ \sigma^2=\ln(1+b^2{\cal M}^2) \]
where $b$ is the turbulent forcing parameter \citep{FKS_08, Fed_ea_10} and ${\cal M}$ is the Mach number.

 \item[3)] ~Mass-density power-law relationship:
 \[ \ln\Bigg(\frac{\rho}{\langle \rho \rangle_L}\Bigg)=x\,\ln\Bigg(\frac{m}{m_0}\Bigg)~,~~~x<0,~~~x=x(L)\vspace{4pt} \]
 \item[4)] ~Self-similar density scaling as expected for turbulent structures: 
\[ \ln\Bigg(\frac{\rho}{\langle \rho \rangle_L}\Bigg)=\alpha\,\ln\Bigg(\frac{l}{l_0}\Bigg)=\frac{3x}{1-x}\,\ln\Bigg(\frac{l}{l_0}\Bigg)~\] 
 \item[5)] ~`Virial-like' equipartition of energies:\\
$|W|\sim f_{\rm gk}E_{\rm kin}~,~~1\lesssim f_{\rm gk}\lesssim 4$,~ (\texttt{wkin\{1-4\}}) \\
$|W|\sim 2E_{\rm kin} + 2E_{\rm th}$,~ (\texttt{wkin2th2}) \\
$|W|\sim 2E_{\rm kin} + E_{\rm mag}$,~ (\texttt{wkin2mag}) \\ 
   \end{itemize}

\section{Alternative models within the chosen approach}
They are built according to the type of object for which the assumptions of mass-density relationship (3) and of a chosen equipartition of energies (5) hold:\vspace*{9pt}
\begin{itemize}
 \item[$\bullet$] ~Statistical {\it ensemble of clumps} \citep[`ensemble model';][]{DVK_11, DVK_12}: Populations (ensembles) of clumps generated at abstract spatial scale $L$ obey a mass-density relationship with power-law index $x(L)$, derived from equipartition relation for their representative members (`typical clumps'). \vspace*{12pt}

 \item[$\bullet$] ~{\it Cloudlet}, defined by a density cut-off (`cloudlet model'): A chosen energy equipartition (assumption 5) determines a density threshold in respect to the mean local density $\langle \rho \rangle_L$. Then the relationship (3) with a power-law index $x(L)$ describes the intrinsic mass-density scaling of structures built into the delineated cloudlet. 
\end{itemize}

\subsection{Clump mass function from the `ensemble model'}

% \figr{Fig/OriB_CMF_poster.eps}{Predictions of the `ensemble model' applied to Orion B: {\it(left)} Model (line) and estimates from dust extinction of the complex substructure in terms of $x(L)$; {\it(right)} Modelled and observational (CO maps) clump mass function (ClMF). The mass range corresponds to spatial scales of clump generation $0.7\lesssim L\lesssim10$~pc (shaded areas).}{12cm}{14cm}{fig_OriB_CMF}

\begin{figure}[!ht]
\begin{center}
\includegraphics[angle=0, width=0.7\textwidth]{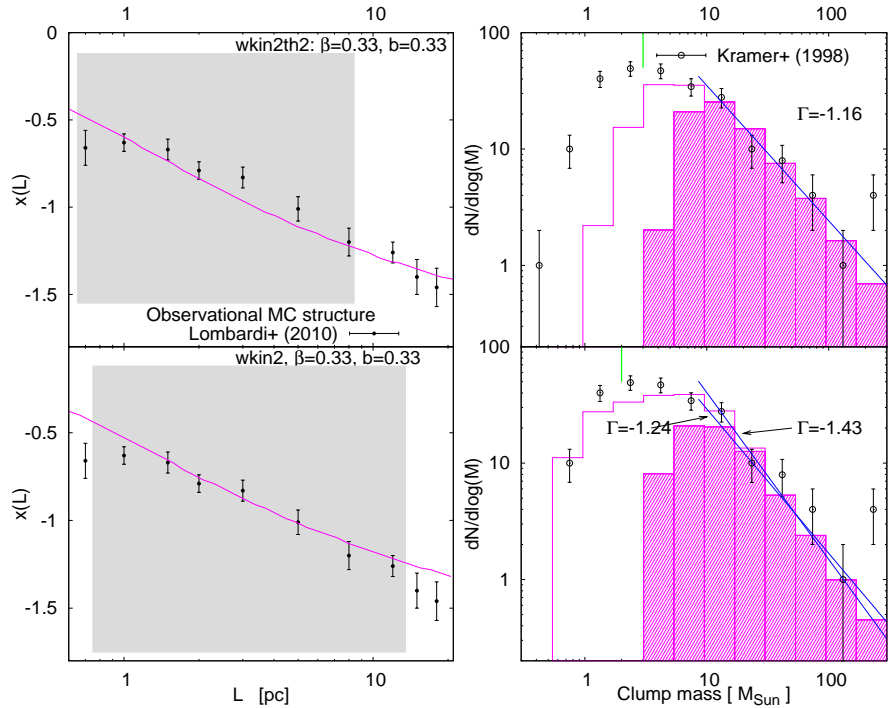}
\caption{Predictions of the `ensemble model' applied to Orion B: {\it(left)} Model (line) and estimates from dust extinction of the complex substructure in terms of $x(L)$; {\it(right)} Modelled and observational \citep[CO maps;][]{Kramer_ea_98} clump mass function (ClMF). The mass range corresponds to spatial scales of clump generation $0.7\lesssim L\lesssim10$~pc (shaded areas).}
\label{fig_OriB_CMF}
\end{center}
\end{figure}

\begin{itemize}
\item The assumption of `virial-like' energy equipartition \texttt{wkin2} or \texttt{wkin2th2} yields a description of MC structure in agreement with dust-absorption maps of large clouds \citep{LAL_10}.

\item The modelled high-mass ClMFs exhibit power-law slopes $\Gamma\sim-1$, as expected for fractal clouds \citep{Elme_97}. When only gravitationally unstable clumps are considered, the slope of their time-weighted high-mass mass function is similar to that of the stellar IMF: $\Gamma\sim-1.3$.
\end{itemize}

\begin{figure}[!ht]
\begin{center}
\includegraphics[angle=0, width=0.55\textwidth]{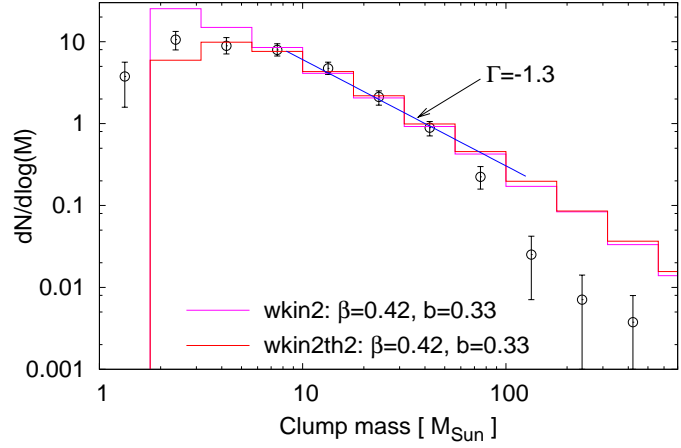}
\caption{Modelled ClMF from various equipartition relations, compared with the observational one, derived for a sample of 9 Galactic MCs \citep{Tachi_ea_02}.}
\label{fig_MCs_CMF}
\end{center}
\end{figure}

\subsection{Clump mass function from the `cloudlet model'}

\begin{itemize}
\item The intrinsic mass-density scaling within the cloudlets is consistent with that of MC fragments delineated by the dendrogram technique in some Galactic clouds \citep{Kauf_ea_10}.

\item Further insights into physics of the dendrogram objects would allow for derivation of the ClMF within the `cloudlet model'.
\end{itemize}

{\it Acknowledgement:} T.V. acknowledges support by the {\em Deutsche Forschungsgemeinschaft} (DFG) under grant KL 1358/15-1.  

\begin{figure}[!ht]
\begin{center}
\includegraphics[angle=0, width=0.85\textwidth]{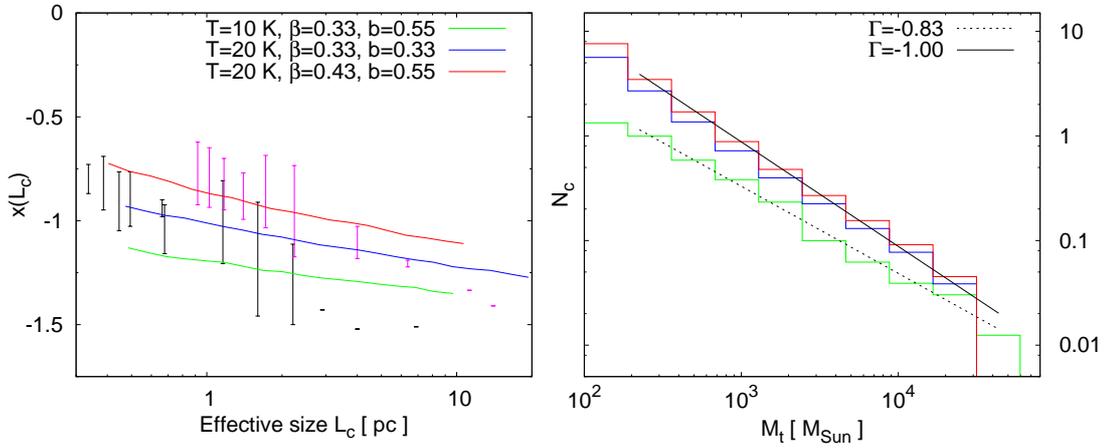}
\caption{Predictions of the `cloudlet model': {\it(left)} Calculated $x(L_{c})$ in cloudlets from the equipartition \texttt{wkin2th2}, compared with observational ranges for MC fragments in Ophiuchus (black) and Perseus (violet), derived from the results of \citet[][see their Fig. 5]{Kauf_ea_10}; {\it(right)} Cloudlet mass functions, derived for the same sets of model parameters. The slopes vary from a typical value for CO clumps to that for fractal clouds.}
\label{fig_MCs_structure}
\end{center}
\end{figure}

\end{document}